\documentclass[12pt]{article}
\usepackage[a4paper]{geometry}
\usepackage[italian,english]{babel}

\usepackage{amsmath}
\usepackage{amsfonts}
\usepackage{amstext}
\usepackage{amssymb}
\usepackage{amsthm}
\usepackage{amscd}

\usepackage[pagebackref,draft=false]{hyperref}
\hypersetup{colorlinks,
linkcolor=myrefcolor,
citecolor=mycitecolor,
urlcolor=myurlcolor}

\usepackage[capitalize]{cleveref}
\usepackage{caption}
\usepackage{etaremune}

\usepackage{xcolor}
\definecolor{myurlcolor}{rgb}{0,0,0.4}
\definecolor{mycitecolor}{rgb}{0,0.5,0}
\definecolor{myrefcolor}{rgb}{0.5,0,0}
\usepackage{graphicx}
\usepackage{tikz}
\usepackage{tikz-cd}
\usepackage{mathrsfs}

\usepackage{etoolbox}
\usepackage{makeidx}
\usepackage{sectsty}
\usepackage{dsfont}
\usepackage{enumitem} 
\usepackage[]{latexsym}
\usepackage{braket}
\usepackage{caption}
\usepackage[utf8]{inputenx}
\usepackage[T1]{fontenc}
\usepackage{lmodern}
\usepackage{textcomp}
\usepackage{microtype}
\usepackage{totcount}
\usepackage{blindtext}

\newtheorem{remark}{Remark}

\newtheorem{example}{Example}

\newtheorem*{proof*}{Proof}

\newcommand{\be}{\begin{equation}}
\newcommand{\ee}{\end{equation}}
\newcommand{\bea}{\begin{eqnarray}}
\newcommand{\eea}{\end{eqnarray}}

\newcommand{\lag}{\mathfrak{L}}

\title{From Classical Trajectories to Quantum Commutation Relations}

\author{F. M. Ciaglia$^{1,5}$  \href{https://orcid.org/0000-0002-8987-1181}{\includegraphics[scale=0.7]{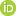}}, G. Marmo$^{2,3,6}$ \href{https://orcid.org/0000-0003-2662-2193}{\includegraphics[scale=0.7]{ORCID.png}}, L. Schiavone$^{4,7}$ \href{https://orcid.org/0000-0002-1817-5752}{\includegraphics[scale=0.7]{ORCID.png}}\\
\footnotesize{$^{1}$\textit{ Max Planck Institute for Mathematics in the Sciences, Leipzig, Germany}} \\
\footnotesize{$^{2}$\textit{ INFN-Sezione di Napoli, Naples, Italy}} \\
\footnotesize{$^{3}$\textit{ Dipartimento di Fisica ``E. Pancini'', Universit\`a di Napoli Federico II,  Naples, Italy}} \\
\footnotesize{$^{4}$\textit{ Department of Mathematics, Faculty of Science, University of Ostrava, Ostrava, Czech Republic}} \\
\footnotesize{$^{5}$\textit{ e-mail: \texttt{florio.m.ciaglia[at]gmail.com}}} \\
\footnotesize{$^{6}$\textit{ e-mail: \texttt{marmo[at]na.infn.it}}} \\
\footnotesize{$^{7}$\textit{ e-mail: \texttt{lucaschiavone[at]live.it}}}
}

\begin{document}

\maketitle

\begin{abstract}
 In describing a dynamical system, the greatest part of the work for a theoretician is to translate experimental data into differential equations. It is desirable for such differential equations to admit a Lagrangian and/or an Hamiltonian description because of the Noether theorem and because they are the starting point for the quantization. As a matter of fact many ambiguities arise in each step of such a reconstruction which must be solved by the ingenuity of the theoretician. In the present work we describe geometric structures emerging in Lagrangian, Hamiltonian and Quantum description of a dynamical system underlining how many of them are not really fixed only by the trajectories observed by the experimentalist. 
\end{abstract}

\tableofcontents

\section{Introduction}

When dealing with the description of a dynamical system in terms of differential equations,if we want to avoid the dry kind of approach: ``Let $M$ be a smooth manifold, and $\Gamma$ a vector field on $M$'', we should consider where this manifold comes from, how a vector field happens to exist on it and what is the relation of this vector field with the observation of an experimentalist. This analysis is an essential part of the work of a theoretician.
Interestingly enough, this point of view is already contained in one of Aristotele's observations (\cite{Aristotele}):
\begin{quotation}
Now, the path of investigation must lie from what is more intimately cognizable and clear to us, to what is clearer and more intimately cognizable in its own nature [...] \\
So we must advance from the concrete data when we have analysed them [...] \\
So we must advance from the concrete whole to the several constituents which it embraces [...]
\end{quotation}
Accordingly, we want to start with ``what is more intimately cognizable and clear to us'' in order to describe the dynamics of a generic dynamical system, that is, we want to start from a set of trajectories on some configuration space $\mathcal{Q}$ extracted from experimental data. 
We avoid here an epistemological digression about the construction of the configuration space from experimental data, and we refer to (\cite{MarmoSaletan}) for a detailed discussion.

In section \ref{Sec:Trajectories} we recall the main points of the construction of differential equations (in general implicit) from the set of  experimental trajectories.
In doing so, we shall see how the carrier manifold for the differential equation is something we should actively build out of experimental data, and not something that is there from the beginning.
Then, assuming the resulting differential equation to be an explicit differential equation\footnote{Some considerations on what happens when we obtain implicit differential equations are made in appendix \ref{App:InverseProblemImplicit}.}, we pass to analyse the process of "tensorialization" of the Lagrangian and Hamiltonian description of the dynamics.
Specifically, in section \ref{sec: lagrangian} we consider the Lagrangian picture of classical mechanics, while in section \ref{sec: hamiltonian} we deal with the Hamiltonian picture.
The tensorial characterization  of the Lagrangian and Hamitlonian pictures allows us to clearly recognize the possibility of alternative Lagrangian and Hamiltonian descriptions for the same dynamical systems.
Furthermore, we comment on the possibility of exploiting alternative tangent bundle structures on the same carrier manifold in order to obtain a Lagrangian descritpion for dynamical systems that are not of second order with respect to a given tangent bundle structure.
This could be particularly relevant in all of those situations in which it is necessary to reparametrize a given dynamics in order to obtain a complete vector field (e.g., the Kepler problem, see \cite{DavanzoKepler, DavanzoQuantum}).
In section \ref{sec: quantum}, we recall how the existence of alternative, nonlinearly related symplectic vector space structures on the same set allows for the definition of alternative and nonlinearly related quantum descriptions in terms of Weyl systems (an important example is provided by the so called \textsc{f}-\textsc{oscillators}, \cite{MankoFoscillator}).

\section{Differential equations from experimental data} \label{Sec:Trajectories}

We model the set $\mathcal{S}$ of observed trajectories as the set of curves over some configuration space subject to suitable regularity conditions:
\be
\mathcal{S}\,=\, \left\{ \gamma \;|\;\; \mathbb{I} \to \mathcal{Q} \;: \;\; t \mapsto \gamma(t) \right\}\,.
\ee
In general, curves in $\mathcal{S}$ intersect each others and we need to discriminate between intersecting curves if we want them to arise as solutions of first order differential equations.
Essentially, we want to "lift" our curves from $\mathcal{Q}$ to a larger manifold where curves do not intersect anymore and may, possibly, be described as solutions of first order differential equations. 

According to the results in  \cite{MarmoSaletan}, if we restrict our attention to Newtonian-like systems, that is, dynamical systems described by second order differential equations, on the configuration space, it is in general sufficient to lift the trajectories of the system from $\mathcal{Q}$ to the tangent bundle $\textbf{T}\mathcal{Q}$ to separate them.
If this is not the case, that is, if trajectories lifted to the tangent bundle still intersect each other, successive lifting are not allowed because they will produce equations of motions of order greater than $2$. Thus, in such a situation, one has to look for other ways to separate trajectories (\cite{MarmoSaletan, MasterThesis}). 
More likely, trajectories furnished by the experimentalist belong to different dynamical systems, for instance they could be particles  with different mass, different charge, different spin or some other characteristic property.

Let us limit our attention to the situation where the set
\be
t\mathcal{S}\,=\, \left\{ t\gamma \;|\;\; \mathbb{I} \to \textbf{T}\mathcal{Q} \;:\;\; t\mapsto \left(\gamma(t), \dot{\gamma}(t)\right) \right\}
\ee
is a set of non-intersecting trajectories on $\textbf{T}\mathcal{Q}$.
It is worth noting that, in general, the set $t\mathcal{S}$ may not coincide with the whole $\textbf{T}\mathcal{Q}$. In any case, it will be required to define a submanifold instead of just a subset of $\textbf{T}\mathcal{Q}$, if it does not coincide with the whole $\textsc{T}\mathcal{Q}$,
\be
t\mathcal{S}\,=\,\mathcal{M} \subset \textbf{T}\mathcal{Q}\,.
\ee
In order to find the differential equation, it is possible to construct the set
\be \label{Eq:SecondLift}
tt\mathcal{S}\,=\, \left\{ tt\gamma \;|\;\; \mathbb{I} \to \textbf{TT}\mathcal{Q} \;:\;\; t \mapsto \left(\gamma(t), \dot{\gamma}(t), \dot{\gamma}(t), \ddot{\gamma}(t)\right) \right\}\,.
\ee
We require again that we are dealing with a submanifold of $\textbf{TT}\mathcal{Q}$
\be
tt\mathcal{S}\,=\,\Sigma \subset \textbf{TT}\mathcal{Q}\,,
\ee
and, when some regularity conditions are satisfied (\cite{MarmoSaletan} chapter $6$), it may happen that $\Sigma$ is the image of a second order vector field
\be
\Gamma \; | \;\; \textbf{T}\mathcal{Q} \to \Gamma\left(\textbf{TT}\mathcal{Q}\right)\,.
\ee
This means that $\dot{\gamma}(t)$ and $\ddot{\gamma}(t)$ in equation \eqref{Eq:SecondLift} are given by the components of a second order vector field over $\textbf{T}\mathcal{Q}$.
In this case it is possible to write the equations of motion for the dynamical system under investigation in the explicit form:
\be \label{Eq:ExplicitEOM}
\dot{\gamma}(t)= \mathbf{v} \qquad \ddot{\gamma}(t) = f(\mathbf{q},\, \mathbf{v})\,,
\ee
where $(\mathbf{q},\, \mathbf{v})$ are coordinates on $\textbf{T}\mathcal{Q}$.

\begin{remark}
There are situations in which the submanifold $\Sigma \subset \textbf{TT}\mathcal{Q}$ can not be written as the image of a second order vector field over $\textbf{T}\mathcal{Q}$. 
Then, $\Sigma \subset \textbf{TT}\mathcal{Q}$ may be interpreted as an  implicit differential equation over $\textbf{T}\mathcal{Q}$ (\cite{MarmoMendellaTulczyjewImplicit} section $5$).
It may also happen that by the lifting procedure we do not cover all of $\textbf{T}\mathcal{Q}$ but only a subset, in this case we would say that our system is subject to some non-holonomic constraints. This would be the case, for instance, when we describe a relativistic particle with $\mathcal{Q}$ being the space-time.
\end{remark}

In physics it is found convenient to ask for a description of the equations of motion in terms of some suitable function, both to be able to exploit Noether's theorem and to deal with quantization procedures. 
In this regard, two possibilities have proven to be quite satisfactory: the \textsc{Lagrangian} and the \textsc{Hamiltonian} pictures. 
In the former case, the function, also called the \textsc{Lagrangian function}, is a real-valued function $\lag \;| \;\; \textbf{T}\mathcal{Q} \to \mathbb{R}$ on the tangent bundle of the configuration space $Q$ giving rise to the following differential equations on $\textbf{T}\mathcal{Q}$, the so called \textsc{Euler-Lagrange equations}:
\be \label{Eq:EulerLagrange}
\begin{split}
\frac{d}{dt} \frac{\partial \lag}{\partial v^j} - \frac{\partial \lag}{\partial q^j} = 0\, \\
\frac{d}{dt}q^j \,- \, v^j \,=\,0 \,. 
\end{split}
\ee
They are clearly implicit differential equations. Indeed, by expanding the time derivatives, one has:
\be
\frac{\partial^2 \lag}{\partial v^i \partial v^j} \, \dot{v}^j + \frac{\partial^2 \lag }{\partial v^i \partial q^j} \, v^j - \frac{\partial \lag}{\partial q^i} \,=\,0
\ee
with $v^i\,=\, \frac{dq^i}{dt}$, where it is clear that the equation may be reduced to an explicit one only if the matrix $\frac{\partial^2 \lag}{\partial v^i \partial v^j}$ is invertible, that is, if the Lagrangian is non-degenerate.
From the geometrical point of view, the specific form of these equations requires the carrier space to possess a tangent bundle structure\footnote{Other approaches, that are useful for dealing with explicitly time-dependent systems, take as fundamental structure the first order jet bundle of the Cartesian product $\mathbb{R}\times \mathcal{Q}$, $\textbf{J}^1 \left( \mathbb{R}\times \mathcal{Q} \right)$, and as Lagrangian, an horizontal density over such a fiber bundle (\cite{KrupkaGlobal} section 4.1, \cite{KrupkovaOVE} chapter $3$ and $4$ and \cite{SardanashvilyNewMethods} chapter 3, pp. 97).}.

In the latter case the function is a real-valued function on the cotangent bundle $\mathscr{H} \;|\;\; \textbf{T}^*\mathcal{Q} \to \mathbb{R}$, also called the \textsc{Hamiltonian function}, giving rise to \textsc{Hamilton equations}:
\be
\frac{d}{dt}q^i \,=\, -\frac{\partial \mathscr{H}}{\partial p_i}  \qquad \frac{d}{dt}p_i \,=\, \frac{\partial \mathscr{H}}{\partial q^i} \,.
\ee
Quite clearly, these equations are always explicit differential equations, independently of the form and properties of $\mathscr{H}$.
As we will explain later in this section, in this Hamiltonian case, the fundamental geometrical structure is the so called \textsc{Liouville one form}, $\theta$, which defines the partial linear structure of the cotangent bundle in a canonical way (\cite{GeometryFromDynamics}) and whose differential is the canonical symplectic form $\omega$ on the cotangent bundle $\textbf{T}^*\mathcal{Q}$, that is, a closed, non-degenerate $2$-form on $\textbf{T}^*\mathcal{Q}$.
In a coordinates system $(\mathbf{q},\,\mathbf{p})$ adapted to the cotangent bundle structure of $\textbf{T}^*\mathcal{Q}$, the \textsc{Liouville one form} reads $\theta=p_{j}\mathrm{d}q^{j}$, while the symplectic form reads $\omega=\mathrm{d}p_{j}\wedge\mathrm{d}q^{j}$ and the partial linear structure is $\Delta \,=\, p_i \, \frac{\partial}{\partial p_i}$.

\begin{remark}
The problem of \, finding a Lagrangian or an Hamiltonian and a Poisson structure such that the equations of motion found in the previous section by means of experimental data coincide with Euler-Lagrange or Hamilton equations is a non-trivial problem in Mathematics. It is, in fact, probably one of the most important problems in the Calculus of Variations and it is the so called \textsc{Inverse problem of the Calculus of Variations}. 

Regarding the Lagrangian formalism, in the "lucky case" where equations of motion are explicit differential equations, a (non-unique) solution is known, provided that some conditions, the so-called  \textsc{Helmholtz conditions}, are satisfied \footnote{See \cite{Santilli} for an analytic introduction, \cite{MarmoTangent, CrampinInverse} for a geometrical approach on the tangent bundle and \cite{KrupkaGlobal, KrupkovaOVE, KrupkovaInverse} for an approach on jet bundles and for an introduction to Variational Sequences.}. For fair implicit differential equations, the inverse problem is in general not widely studied (see however \cite{DeDiegoInverse, DeDiegoConstrainedInverse}), and we refer to appendix \ref{App:InverseProblemImplicit} for a discussion about the difficulties in formulating the problem.

In the Hamiltonian case, the problem is slightly more complicated because we should be looking for both a Hamiltonian and a Poisson structure at the same time. The existence of a possible Poisson description, according to Dirac, seems to be a necessary condition to carry on a quantization procedure (\cite{MarmoGiordanoRubanoInverseHamiltonian}).
\end{remark}

\section{Dynamical systems and geometrical structures:\\ Lagrangian picture} \label{sec: lagrangian}

A great boost in our understanding of the Lagrangian and Hamiltonian formalisms mentioned above came when a tensorial characterization of the fundamental geometric structures underlying these descriptions of the dynamics was achieved.
Regarding the Lagrangian formulation of dynamics, a necessary step in the ``tensorialization'' process is a tensorial characterization  of the structure of tangent bundle  and second order (Newtonian-like) vector fields (as given for instance in (\cite{MarmoTensorialTangentBundle})). 
Essentially, given a smooth manifold of dimension $2n$, say $\mathcal{M}$, a tangent bundle structure for $\mathcal{M}$ is encoded in the following two objects:

\begin{itemize}
\item a $(1,\,1)$ tensor field, say $S$, such that $S^2 = 0$, $Ker\,S = Im\,S$ and $\mathcal{N}_S = 0$, where $\mathcal{N}_{S}$ denotes the Nijenhuis tensor associated with $S$ (\cite{MarmoTangent} section $2.4$);
\item a partial linear structure $\Delta$ (see definition $3.15$ at page $158$ in \cite{GeometryFromDynamics}), whose critical points constitute a smooth submanifold of $\mathcal{M}$ of dimension $n$, and such that $\mathcal{L}_{\Delta}\,S = -S$ and $S(\Delta) = 0$.
\end{itemize}

We do not give a proof of the theorem for which we refer to \cite{MarmoTensorialTangentBundle}. However, the condition $Ker\,S = Im\,S$ tells us that $\mathcal{M}$ is even-dimensional and then the whole proof is based on the construction of an atlas over $\mathcal{M}$ in which $S$ and $\Delta$ take, in a local chart of such an atlas, their "canonical" form:
\be
S = \frac{\partial}{\partial v^i} \otimes dq^i \qquad \Delta = v^i \frac{\partial}{\partial v^i}\,,
\ee
where the $q^{j}$'s may be then interpreted as configuration-like coordinates, while the $v^{j}$ may be interpreted as velocity-like coordinates, and we have $\mathcal{M}\cong TQ$. 
Then, it is clear that the only diffeomorphisms preserving the given tangent bundle structure encoded in $\Delta$ and $S$ are all those diffeomorphisms $\Phi\colon \mathcal{M}\rightarrow \mathcal{M}$ that are the tangent lift $T\phi$ of some diffeomorphism $\phi\colon \mathcal{Q}\rightarrow \mathcal{Q}$.
Locally, these diffeomorphisms may always be written as:
\be
q'^i = f^i(\textbf{q}) \qquad v'^i = v^j\,\frac{\partial f^i}{\partial q^j}\,.
\ee
In this sense, $S$ and $\Delta$ represent the tangent bundle structure and identify the subgroup of diffeomorphisms which preserve them both.

In this framework, a second order (Newtonian-like) vector field $\Gamma$ on  $\textbf{T}\mathcal{Q}\cong\mathcal{M}$ is defined as a vector field the integral curves (on $\textbf{T}\mathcal{Q}\cong\mathcal{M}$) of which are tangent lifts of curves on the base manifold $\mathcal{Q}$.
It is easy to see that, in every coordinate system $(\mathbf{q}, \, \mathbf{v})$ which is adapted to the tangent bundle structure on  $\textbf{T}\mathcal{Q}\cong\mathcal{M}$ encoded in $S$ and $\Delta$, the second order (Newtonian-like) vector field $\Gamma$ can be written as:
\be
\Gamma\,=\,v^{j}\,\frac{\partial}{\partial q^{j}} \,+\, \Gamma^{j}(q,v)\,\frac{\partial}{\partial v^{j}} \,.
\ee
From this it is possible to give a purely tensorial description of a second order vector field in terms of $S$ and $\Delta$ as follows
\be \label{Eq:SODECondition}
S(\Gamma) = \Delta\,.
\ee
This tensorial characterization of second order (Newtonian-like) vector fields allows us to immediately see what happens to the second order vector field $\Gamma$ when we perform arbitrary transformations on the manifold $\mathcal{M}\cong \textbf{T}\mathcal{Q}$.
Consequently, let us consider a diffeomorphism $\phi \colon\mathcal{M}\rightarrow\mathcal{M}$ which is not necessarily adapted to the tangent bundle structure on $\mathcal{M}$ encoded in $S$ and $\Delta$.
Since $\phi$ is a diffeomorphism, it is possible to transform $\Gamma$ by means of $\phi$ to obtain another vector field on $\mathcal{M}$.
Then, we may look at what happens to equation \ref{Eq:SODECondition} under the action of $\phi$ obtaining:
\be
\phi_* \left[ S(\Gamma) \right] = \phi_*\Delta \;\; \to \;\; \left(\phi^*S \right)\, \left(\phi_* \Gamma \right) = \phi_* \Delta\,.
\ee
This equation is interpreted as the SODE condition for the "transformed" field, $\phi_*\Gamma$ with respect to the "transformed" tangent bundle structure given by $\phi^*S$ and $\phi_*\Delta$.
From this, it follows that an arbitrary transformation on $\mathcal{M}$ will transform our vector field into a vector field which is still a second order vector field, but, in general, with respect to another alternative tangent bundle structure on $\mathcal{M}$.
In particular, when $\phi$ is a symmetry for $\Gamma$, that is, if $\phi_* \Gamma = \Gamma$, we obtain:
\be
\left( \phi^*S \right) \Gamma = \phi_* \Delta\,,
\ee
which means that the vector field $\Gamma$ itself is a second order vector field also for the alternative tangent bundle structure given by $\phi^*S$ and $\phi_*\Delta$.

The notion of second-order vector field, together with the tensorial characterization of the tangent bundle structure encoded in $S$ and $\Delta$, allows to give a tensorial formulation of the Euler-Lagrange equations. 
Given the Lagrangian function $\lag$, we define the $1$-form:
\be
\theta_{\lag}\,:=\,\mathrm{d}_{S}\,\lag\,,
\ee
where $\mathrm{d}_{S}$ is defined as $\mathrm{d}_S \,:=\, S \circ \mathrm{d}$ according to formula $(2.4.12)$ at page $170$ \cite{MarmoTangent}, and its action on functions reads $S\circ\mathrm{d}$.
The one-form $\theta_{\lag}$ is a semi-basic form on $\textbf{T}\mathcal{Q}$, and its coordinate expression in a coordinates system $(\mathbf{q},\,\, \mathbf{v})$ adapted to the tangent bundle structure of $\textbf{T}\mathcal{Q}$ reads:
\be
\theta_{\lag} = \frac{\partial \lag}{\partial v^i} \, dq^i\,.
\ee
Now, we may introduce the so-called Lagrangian $2$-form:
\be
\omega_{\lag}\,:=\,-\mathrm{d}\,\theta_{\lag}\,.
\ee
This is a closed $2$-form on $\textbf{T}\mathcal{Q}$, and we may consider the equation:
\be\label{Eq:IntrinsicEulerLagrange1}
i_{\Gamma}\,\omega_{\lag}\,=\,\mathrm{d}E_\lag\,,
\ee
where $\Gamma$ is a vector field on $\textbf{T}\mathcal{Q}$ and $E_\lag := \Delta(\lag) - \lag$.
In general, this equation does not admit of a unique solution. Indeed, if the two-form is degenerate it may happen that there is no solution and when a solution exists it need not be unique.
However, if $\omega_\lag$ is non-degenerate\footnote{Nondegenerancy of $\omega_{\lag}$ is equivalent to $\mathrm{det}(\frac{\partial^{2}\lag}{\partial v^{j}\partial v^{k}})\neq 0$ (\cite{MarmoTangent} section $3$). 
When this is not the case, a careful analysis of the given situation is needed since, in general, we are in the presence of the description of a physical system in terms of fair implicit differential equations.}, then it is a symplectic form and equation \eqref{Eq:IntrinsicEulerLagrange1} admits a unique solution $\Gamma$ which is a second-order vector field on $\textbf{T}\mathcal{Q}$.
Using the \textsc{Cartan's identity} (i.e., $\mathcal{L}_{X} = i_X \, \mathrm{d} + \mathrm{d} \, i_X$) it is possible to show that equation \eqref{Eq:IntrinsicEulerLagrange1} is equivalent to 
\be \label{Eq:IntrinsicEulerLagrange}
\mathcal{L}_\Gamma \,\theta_{\lag} - \mathrm{d}\lag = 0\,.
\ee
Furthermore, it is easy to see that, in a coordinates system $(\mathbf{q},\,\, \mathbf{v})$ adapted to the tangent bundle structure of\, $\textbf{T}\mathcal{Q}$, equation \eqref{Eq:IntrinsicEulerLagrange} acquires the usual Euler-Lagrange form:
\be
\frac{d}{dt} \frac{\partial \lag}{\partial v^j} - \frac{\partial \lag}{\partial q^j} = 0\,. 
\ee

Now that we have a tensorial formulation of Euler-Lagrange equations, we may consider some diffeomorphism $\phi$ on $\mathcal{M}\cong\textbf{T}\mathcal{Q}$ and look at the "transformation properties" of  Euler-Lagrange equations.
We will do it by looking at how the differential $1$-form on the left hand side of (\ref{Eq:IntrinsicEulerLagrange}) changes by taking its pull-back trough the diffeomorphism $\phi$. 
By using theorem $7.4.4$ at page $413$ and proposition $7.4.10$ at page $416$ in \cite{abraham_marsden_ratiu-manifolds_tensor_analysis_and_applications}, and by using the \textsc{Cartan's identity} for the Lie derivative, we have
\be
\mathcal{L}_{\phi^* \Gamma} \, \phi^* \theta_{\lag} - \mathrm{d}\, \phi^* \lag = 0
\ee
Again, if $\phi$ is the tangent lift of a diffeomorphism on $\mathcal{Q}$ (i.e., $\phi=T\varphi$ for some $\varphi\colon\mathcal{Q}\rightarrow\mathcal{Q}$), and when it is a symmetry for $\Gamma$ (i.e., $\phi^{*}\Gamma=\Gamma$), we obtain different Lagrangian descriptions for the same dynamical system.
For a complete geometrical characterization of alternative Lagrangian descriptions we refer to (\cite{MarmoTangent}).

\begin{remark}
Until now, we have always started with a vector field which is of second order with respect to some tangent bundle structure on $\mathcal{M}$.
However, we may think of starting with a generic vector field which is not a second order one and ask if and how it is possible to "transform" it into a second order one with respect to some yet undetermined tangent bundle structure on $\mathcal{M}$ (if it exists). 
In particular, given a  vector field $\Gamma$ over the tangent bundle of some configuration space $\mathcal{Q}$, say $\mathcal{M} \cong \textbf{T}\mathcal{Q}$, characterized by $S$ and $\Delta$, we may ask if it is possible to endow $\mathcal{M}$ with an alternative tangent bundle structure which makes the given vector field into a second order one.
We will now write down some simple examples of such a situation in a coordinate-dependent fashion, postponing a tensorial analysis to a future work.

\clearpage

\begin{example}{Example}
Let us consider a configuration space $\mathcal{Q} \cong \mathbb{R}$. Then $\textbf{T}\mathcal{Q} \cong \textbf{T}\mathbb{R} \cong \mathbb{R}^2$. Consider a coordinatization $(q,\,v)$ over $\textbf{T}\mathcal{Q}$ and the vector field:
\be
\Gamma = f(v)v \frac{\partial}{\partial q}
\ee
where $f$ is a nowhere vanishing function and $\mathrm{d}(f(v)\,v) \,\neq\, 0$. This vector field would describe a reparametrized free particle with a reparametrization-function which is a constant of the motion. We may need a more general reparametrization if we want to turn a vector field into a complete one (this would be the case for the Kepler problem. See \cite{DavanzoKepler}). Now, consider a diffeomorphism:
\begin{equation}
\phi \;|\;\; \textbf{T}\mathcal{Q} \to \textbf{T}\mathcal{Q} \;:\;\; (q,\,v) \mapsto \left( \phi_1(q,\,v)=:y ,\, \phi_2(q,\,v)=:w \right)
\end{equation}
then we can transform $\Gamma$ through $\phi^{-1}$:
\begin{equation}
\left(\phi^{-1}\right)_*\Gamma = f(v)v \left( \frac{\partial y}{\partial q}\frac{\partial}{\partial y} + \frac{\partial w}{\partial q} \frac{\partial}{\partial w} \right)
\end{equation} 

Looking at the previous expression, in order to obtain a second order vector field, we need a diffeomorphism such that:
\be
f(v)v \frac{\partial y}{\partial q} = w
\ee
It is easy to see that if we consider the following diffeomorphism:
\be
y = \frac{q}{f(v)} \qquad w = v
\ee
then:
\be
\left(\phi^{-1}\right)_*\Gamma = w \frac{\partial}{\partial y}
\ee
which is a second order vector field with respect to the tangent bundle structure characterized by:
\be
S' = \mathrm{d}y \otimes \frac{\partial}{\partial w} \qquad \Delta' = w \frac{\partial}{\partial w}
\ee
\end{example}

\begin{example}{Example}
Now, let us consider the following vector field on $\textbf{T}\mathbb{R}$:
\be
\Gamma = \omega\,v\frac{\partial}{\partial q} - \omega\, q \frac{\partial}{\partial v}
\ee
where $\omega$ is a constant of the motion for $\Gamma$. This is again a reparametrized vector field with a reparametrization function which is a constant of the motion. Consider, again a diffeomorphism:
\begin{equation}
\phi \;|\;\; \textbf{T}\mathcal{Q} \to \textbf{T}\mathcal{Q} \;:\;\; (q,\,v) \mapsto \left( \phi_1(q,\,v)=:y ,\, \phi_2(q,\,v)=:w \right)
\end{equation}
and let us evaluate the following vector field:
\begin{equation}
\left(\phi^{-1} \right)_* \Gamma = \omega \left( v\frac{\partial y }{\partial q} - q \frac{\partial y}{\partial v} \right) \frac{\partial}{\partial y} + \omega \left( v \frac{\partial w}{\partial q} - x\frac{\partial w}{\partial v} \right)\frac{\partial}{\partial w}
\end{equation}
then it is a matter of direct computation to show that, if one takes:
\begin{equation}
y = \frac{q}{\omega^2} \qquad w = \frac{v}{\omega}
\end{equation}
then:
\begin{equation}
\left(\phi^{-1} \right)_* \Gamma = w \frac{\partial}{\partial y} - \omega^2\, y \frac{\partial}{\partial w}
\end{equation}
which is a second order vector field with respect to the tangent bundle structure given by:
\begin{equation}
S' = \mathrm{d}y \otimes \frac{\partial}{\partial w} \qquad \Delta' = w \frac{\partial}{\partial w}
\end{equation}
and which represent an harmonic oscillator of frequency $\omega$, where the frequency is a constant of the motion, it is an example of \textsc{f-oscillator} (see \cite{MankoFoscillator}). 
\end{example}

\begin{example}{Example}
Now, let us consider the following vector field on $\textbf{T}\mathbb{R}$:
\begin{equation}
\Gamma = q \frac{\partial}{\partial q}
\end{equation}
By means of a diffeomorphism, as in the previous two examples, one has:
\begin{equation}
\left(\phi^{-1} \right)_* \Gamma = q \left( \frac{\partial y}{\partial q} \frac{\partial}{\partial y} + \frac{\partial w}{\partial q} \frac{\partial}{\partial w} \right)
\end{equation}
Here, it is easy to show that if one takes the following diffeomorphism:
\begin{equation}
y = q + f(v) \qquad w = q
\end{equation}
with $\mathrm{d}f \wedge \mathrm{d}q \,\neq \,0$, then:
\be
\left(\phi^{-1} \right)_* \Gamma = w \frac{\partial}{\partial y} + w \frac{\partial}{\partial w}
\ee
which is, again, a second order vector field with respect to the alternative tangent bundle structure characterized by:
\be
S' = \mathrm{d}y \otimes \frac{\partial}{\partial w} \qquad \Delta' = w \frac{\partial}{\partial w}
\ee
\end{example}

\end{remark}

\begin{remark}
Postponing to a future work a carefully analysis of such a question, we would stress the fact that we have considered a generic vector field (not of second order with respect to $S$ and $\Delta$) and we have searched for a diffeomorphism which transforms such a vector field into a second order one with respect to a novel tangent bundle structure. On the other side another situation is possible, that is, the one in which the given vector field is made into a second order one by selecting an appropriate tangent bundle structure. Of course the two situations are strictly related.
\end{remark}

\section{Dynamical systems and geometrical structures:\\ Hamiltonian picture} \label{sec: hamiltonian}

As said before, in the Hamiltonian formalism, the fundamental geometric structure is the Liouville $1$-form, $\theta$\footnote{$\theta = \gamma \circ T\pi$, where $\gamma$ is a section of $\textbf{T}^*\mathcal{Q}$ and $\pi$ is the canonical projection of the cotangent bundle.}, whose differential is the canonical symplectic form $\omega$ on the cotangent bundle $\textbf{T}^*\mathcal{Q}$, that is, a closed, non-degenerate $2$-form on $\textbf{T}^*\mathcal{Q}$.
Once we have $\omega$ and the Hamiltonian function $\mathscr{H}$, we may always write the equation
\be\label{eqn: Hamilton equations}
i_{X_{\mathscr{H}}}\,\omega\,=\,\mathrm{d}\mathscr{H}\,,
\ee
where $X_{\mathscr{H}}$ is a vector field on $\textbf{T}^*\mathcal{Q}$.
Unlike the Lagrangian setting, the $2$-form $\omega$ is always nondegenerate, and thus the previous equation has a unique solution $X_{\mathscr{H}}$ which is called the Hamiltonian vector field associated with $\mathscr{H}$.
It is immediate to check that $X_{\mathscr{H}}$ is such that $\mathcal{L}_{X_{\mathscr{H}}}\omega=0$, and that, in a coordinates system $(\mathbf{q},\,\mathbf{p})$ adapted to the cotangent bundle structure of $\textbf{T}^*\mathcal{Q}$, equation \eqref{eqn: Hamilton equations} "becomes" Hamilton equations:
\be
\frac{d}{dt}q^i \,=\, \frac{\partial \mathscr{H}}{\partial p_i} =  \{q^i,\, \mathscr{H}\} \qquad \frac{d}{dt}p_i \,=\, -\frac{\partial \mathscr{H}}{\partial q^i} = -\{p_i,\, \mathscr{H}\}\,.
\ee
Furthermore, it is important to note that $\omega$ is independent of the Hamiltonian function $\mathscr{H}$. It has a "kinematical" character, it is canonically defined on any cotangent bundle independently of any possible dynamics.
This is clearly in contrast with what happens in the Lagrangian setting where the Lagrangian $2$-form $\omega_{\lag}$ always depends on the Lagrangian function $\lag$. In some sense $\lag$ has both  "kinematical" and  "dynamical" contents, while in the Hamiltonian picture the symplectic structure has a kind of "universal" character.

Since $\omega$ is a symplectic form, it is invertible, and its inverse is denoted by $\Lambda$. We recall that $\omega$ defines a base-invariant fiberwise isomorphism between $\textbf{TT}^*\mathcal{Q}$ and $\textbf{T}^*\textbf{T}^*\mathcal{Q}$:
\be
\omega^{\flat}\;\;|\;\; \textbf{TT}^*\mathcal{Q} \to \textbf{T}^*\textbf{T}^*\mathcal{Q} \;\;:\;\; (\mathbf{q},\, \mathbf{X})\mapsto (\mathbf{q},\, i_{\mathbf{X}}\omega)
\ee
then $\Lambda$ is the inverse of this isomorphism.
In a coordinates system $(\mathbf{q},\,\mathbf{p})$ adapted to the cotangent bundle structure of $\textbf{T}^*\mathcal{Q}$, we have $\Lambda = \frac{\partial}{\partial q^i} \wedge \frac{\partial}{\partial p_i}$.
The bivector field $\Lambda$ allows us to define a \textsc{Poisson bracket} (See \cite{GeometryFromDynamics} section $4.3.1$) on smooth functions on $\textbf{T}^*\mathcal{Q}$ in the following way:
\be
\{f,\,g\} = \Lambda(df,\,dg) \;\; \forall \,f,\, g \in \mathcal{C}^{\infty}(\textbf{T}^*\mathcal{Q})\,.
\ee
Clearly, once we have $\Lambda$ and $\mathscr{H}$ it is immediate to check that equation \eqref{eqn: Hamilton equations} may be alternatively written as
\be\label{eqn: Poisson-Hamilton equations}
X_{\mathscr{H}}\,=\,i_{\mathrm{d}\mathscr{H}}\,\Lambda\,.
\ee
It is clear that in general one may use a bi-vector field $\Lambda$ which is not invertible and previous equations would still make sense.
Indeed, Hamilton equations may be formulated starting only with a Poisson Bracket, that is, a bivector field $\Lambda$ satisfying $[\Lambda,\,\Lambda]=0$, where $[\, \cdot \, , \,\cdot \,]$ denotes the  Schouten – Nijenhuis bracket (See Appendix $D$ in \cite{GeometryFromDynamics}).
If $\Lambda$ is the inverse of some symplectic form $\omega$, then $[\Lambda,\,\Lambda]=0$ is automatically satisfied, but not every Poisson tensor comes from a symplectic form.
Indeed, it would be possible to have  a Poisson bi-vector field which is not invertible, and thus, which is not the inverse of a symplectic form.
This is the case, for example, of the canonical Poisson bi-vector field on the dual of the Lie algebra of a Lie group. 
As an example, consider the Lie algebra $\mathfrak{g}$ of a finite-dimensional Lie group $G$, and the dual $\mathfrak{g}^{*}$ of $\mathfrak{g}$.
The elements in $\mathfrak{g}$ may be identified with the linear functions on the vector space $\mathfrak{g}^{*}$ by means of the map $a\mapsto f_{a}$ where $a$ is in $\mathfrak{g}$ and $f_{a}$ is the linear function on $\mathfrak{g}^{*}$ given by
\be
f_{a}(\xi)\,=\,\xi(a)
\ee
for every $\xi\in\mathfrak{g}$.
Since the differentials of the linear functions on the vector space $\mathfrak{g}^{*}$ generate the module of differential one-forms on $\mathfrak{g}^{*}$, we may define a bi-vector field $\Lambda$ by
\be
\Lambda(\mathrm{d}f_{a},\,\mathrm{d}f_{b})\,:=\,f_{[a,\,b]}\,,
\ee
where $[\, ,\,]$ denotes the Lie product in $\mathfrak{g}$, and extending $\Lambda$ by linearity.
Then, the fact that the bi-vector field $\Lambda$ satisfies $[\Lambda,\,\Lambda]=0$ essentially follows from the fact that $[\, ,\,]$ satisfies the Jacobi identity (see \cite{GeometryFromDynamics, PichereauPoisson}).
Furthermore, it is easy to see that $\Lambda$ is in general not invertible (for instance when $\mathfrak{g}$ is the Lie algebra of the unitary group $\mathcal{U}(\mathcal{H})$ of some finite-dimensional complex Hilbert space $\mathcal{H}$).

Note that equation \eqref{eqn: Hamilton equations} is well-defined not only on the cotangent bundle $\textbf{T}^*\mathcal{Q}$ of some configuration space $\mathcal{Q}$, but may be defined on any symplectic manifold in the sense that if $\mathcal{M}$ is any $2n$-dimensional smooth manifold endowed with a symplectic form $\omega$, then, it always makes sense to define the Hamiltonian vector field associated with a given Hamiltonian function according to equation \eqref{eqn: Hamilton equations}, for instance, on the $2$-dimensional sphere.
Furthermore, equation \eqref{eqn: Poisson-Hamilton equations}, as we have already remarked, makes sense in an even more general context because $\mathcal{M}$ needs not be even-dimensional in order for a Poisson bivector $\Lambda$ to exists on it.

Now, let us consider a Hamiltonian vector field  system with respect to the symplectic structure given by $\Omega$. Let us consider a diffeomorphism $\phi \;|\;\; \mathcal{M} \to \mathcal{M}$.
It is immediate to check that \eqref{eqn: Hamilton equations} changes equivariantly in the sense that $\phi^{*}\Omega$ is another symplectic form on $\mathcal{M}$, and $\phi^{*}X_{\mathscr{H}}$ is the Hamiltonian vector field associated with $\phi^{*}\mathscr{H}$ by means of $\phi^{*}\Omega$.
If $\phi$ is a symmetry for our Hamiltonian vector field (that is, $\phi^* X_{\mathscr{H}} = X_{\mathscr{H}}$), then, we set $\Omega_{\phi} := \phi^* \Omega$, and we define the $(1,\,1)$ tensor field associated with $\phi$ and the initial symplectic form
\be
T_{\phi} = \Lambda \circ \Omega_\phi\,.
\ee
Since $\Omega$ and $\Omega_{\phi}$ are invariant with respect to $X_{\mathscr{H}}$, it follows that $T_{\phi}$ is also invariant.
Then, according to \cite{MarmoTangent} (formula $(2.4.12)$ at page $170$), it is possible to define the $T$-differential, say $\mathrm{d}_T$ which acts on functions as $T \circ \mathrm{d}$. 
Now, given a smooth constant of the motion for our Hamiltonian field, that is, a smooth function $f$ such that $\mathcal{L}_{X_{\mathscr{H}}} f = 0$, it is possible to define the following closed $2$-form on $\mathcal{M}$
\be
\omega_f = \mathrm{d} \mathrm{d}_T f\,.
\ee
It is immediate to check that $\omega_{f}$ is invariant with respect to $X_{\mathscr{X}}$, indeed
\be
\begin{split}
\mathcal{L}_{X_{\mathscr{H}}} \omega_f & = \mathcal{L}_{X_{\mathscr{H}}}\left(\mathrm{d} \mathrm{d}_T f\right)\,=\,\mathrm{d}\left(\mathcal{L}_{X_{\mathscr{H}}}\left(T\circ\mathrm{d} f\right)\right)\,= \\ 
\,&=\,\mathrm{d}\left(\left(\mathcal{L}_{X_{\mathscr{H}}}T\right)\circ\mathrm{d} f\right) +\mathrm{d}\left(T\circ\mathcal{L}_{X_{\mathscr{H}}}\left(\mathrm{d} f\right)\right)\,=\,0\,.
\end{split}
\ee
If $\omega_{f}$ is nondegenerate (i.e., symplectic), this is equivalent to $i_{X_{\mathscr{H}}} \omega_f$ being a non-zero closed $1$-form.
It may happen that $i_{X_{\mathscr{H}}} \omega_f$ is actually exact (for instance if $\mathcal{M}$ is contractible), in which case we obtain another Hamiltonian description of $X_{\mathscr{H}}$ with respect to another symplectic form and another Hamiltonian function.
It is worth noting that this procedure may be iterated, that is, out of any constant of the motion one may construct an invariant $2$-form which, if non-degenerate, is a new, alternative, symplectic form. Moreover, consider two of such symplectic forms, say $\omega_1$ and $\omega_2$, and consider the two associated Poisson braket. If their sum is again a Poisson braket, then the two symplectic structure are \textsc{compatible in the sense of Magri}\footnote{See \cite{SmirnovMagriCompatibility}} that is, the recursion operator $N \,=\, \omega_2^{\sharp} \, \omega_1^{\flat}$ may have $n$ simple eigenvalues which may turn out to be $n$ functionally independent constant of the motion in involution and thus, by \textsc{Arnold-Liouville theorem}, the Hamiltonian, with respect both the symplectic structures, field is also completely integrable. 

To resume, starting with a symplectic structure, an Hamiltonian and a symmetry for the corresponding Hamiltonian vector field which is not a canonical symmetry, it is possible to construct, in principle, other symplectic forms that provide an alternative description of the same Hamiltonian field. Moreover, under suitable conditions, they are compatible in the sense of Magri and thus guarantee the complete integrability of the Hamiltonian system (See also \cite{SmirnovMagriCompatibility}). 

\section{Dynamical systems and geometrical structures:\\ Quantum systems} \label{sec: quantum}

The existence of alternative symplectic structures invariant under the infinitesimal action of a dynamical vector field allows for the existence of alternative quantum descriptions as it may be shown by means of the Weyl formalism.
The subject of this section is to briefly review how these alternative quantum descriptions arise.
For this purpose, we need to recall what \textsc{Weyl systems} are. 
Let us consider a symplectic vector space $(\mathbb{V},\,\omega)$, and a complex, separable Hilbert space say $\mathcal{H}$. 
A \textsc{Weyl system} is a map from $\mathbb{V}$ to the group $\mathcal{U}(\mathcal{H})$ of unitary operators over $\mathcal{H}$
\be
W \;|\;\; \mathbb{V} \to \mathcal{U}(\mathcal{H}) \; : \;\; z \mapsto W(z) 
\ee
such that
\be \label{Eq:WeylCCR1}
W(z\,+\,z')\,=\,W(z)W(z')\, e^{-\frac{i}{2\hbar} \omega(z,\,z')}\,,
\ee 
so that 
\be \label{Eq:WeylCCR}
W(z)W(z') \,=\, W(z')W(z) \, e^{\frac{i}{\hbar} \omega(z,\,z)} \,,
\ee
which means that the operators associated with elements in the same Lagrangian subspace\footnote{We recall that given a symplectic vector space, say $(\mathbb{V},\,\omega)$, a Lagrangian subspace of $\mathbb{V}$ with respect to the symplectic structure $\omega$, say $\mathbb{L}$, is a subspace $\mathbb{L} \,=\, \{ z \in \mathbb{V} \;:\;\; \omega(z_i ,\, z_j)\,=\,0 \;\; \forall z_i,\,z_j \in \mathbb{L} \}$.} of $(\mathbb{V},\, \omega)$ commute.
Furthermore, $W$ is required to be \textsc{strongly continuous}, that is, it must hold that
\be
lim_{z \to z_0} \, ||W(z) \,-\, W(z_0)|| _{sup} = 0 \;\; 
\ee
where $|| \,\cdot\,||_{sup}$ is the $sup$ norm on the Banach space of linear operators over $\mathcal{H}$, $\mathcal{OP}(\mathcal{H})$.

From equation \eqref{Eq:WeylCCR1}, it follows that when $z$ and $z'$ are in the same one-dimensional subspace of $\mathbb{V}$, that is, $z'=az$ for some $a\in\mathbb{R}$, then:
\be
W(z + z') \,=\, W(z)\,W(z')\,,
\ee
that is, $W(az)$ is a one-parameter group of unitary operators labelled by the parameter $a \in \mathbb{R}$.
Being $W$ strongly continuous, the \textsc{Stone-Von Neumann theorem} implies that 
\be
W(az) \,=\, e^{ia \,G(z)}
\ee
for some (possibly unbounded) self-adjoint operator $G$. 
Thus,  equation \eqref{Eq:WeylCCR} may be written as
\be
e^{i \,G(z)} e^{i \,G(z')} \, e^{-i \,G(z)} e^{-i \,G(z')}  \,=\,  e^{\frac{i}{\hbar} \,\omega(z,\,z')}\,,
\ee
from which we recognize \textsc{Weyl commutation relations}. 
If $\left\{ \phi_{t} \right\} _{t\in\mathbb{R}}$ is a
one-parameter group of symplectomorphisms,
then we can define:
\begin{equation}
W_{t}\left( z\right) =:\mathcal{W}\left( \phi_{t}(z)\right).
\end{equation}
This is a one parameter group of unitary transformations which may be represented as a similarity transformation by means of a one-parameter group of automorphisms on the space of operators, namely $
\left\{ \phi_{t}\right\} $:
\begin{equation}\label{eqn: quantum Hamiltonian evolution in Weyl systems}
W_{t}\left( z\right) =e^{it\widehat{H}}\,W\left( z\right)\,
e^{-it\widehat{H}}\,,
\end{equation}
where $\widehat{H}$ is the infinitesimal generator derived by Stone-Von Neumann theorem.
In some cases,  $\phi_{t} $ may arise as the flow associated with the dynamical evolution generated by a linear vector field $\Gamma$ on $\mathbb{V}$.

At this point, it is impossible not to recall von Neumann's theorem.
This theorem states that Weyl systems do exist for any
finite-dimensional symplectic vector space $\mathbb{V}$, and provide an explicit realization  for both $W$ and the Hilbert space $\mathcal{H}$.
Specifically, $\mathbb{V}$ is decomposed (in a not unique way) into the direct sum of Lagrangian subspaces\footnote{Given an even dimensional symplectic vector space, such a decomposition is always possible (\cite{GeometryFromDynamics} section $5.2.2$).}, $\mathbb{V} \,=\, \mathbb{L}_1 \, \oplus \, \mathbb{L}_2$, and we define $U=:W|_{\mathbb{L}_2}$ and $V=:W|_{\mathbb{L}_1}$.
Then, the Hilbert space $\mathcal{H}$ is realized as the space $\mathcal{H}=L^{2}\left(
L,d^{n}x\right) $  of square-integrable complex
functions with respect to the translationally-invariant Lebesgue measure on the Lagrangian subspace $\mathbb{L}_1$.
Finally, the Weyl system is realized as follows:

\begin{eqnarray} \label{Eq:WeylSystem}
&&\left(V\left( z_1\right) \psi \right) \left( x\right) =\psi
\left( x+z_{1}\right) \\
&&\left( U\left( z_{2} \right) \psi \right) \left( x\right)
=e^{i\,\omega\left( x,z_{2} \right) }\psi \left( x\right)\,.  \notag
\end{eqnarray}
where $z_{1},x\in \mathbb{L}_1$ and $z_{2} \in \mathbb{L}_2$.
Setting $z_{1}\equiv q$ and $z_{2}\equiv p$, the infinitesimal  generators of $U(z_1)$ and $V(z_2)$ are:
\be
\begin{split}
V(z_1)\,&=\, e^{i q^j P_j} \, \rightarrow \, \left(P_j \,\psi\right)(\textbf{x}) \,= i\frac{d}{dx^j} \psi(\textbf{x})\,\\
U(z_2)\,&=\, e^{i p_j Q^j} \, \rightarrow \, \left(Q^j \,\psi\right)(\textbf{x}) \,= q^j \psi(\textbf{x}) \,.
\end{split}
\ee
The definition of Weyl system, as well as its explicit realization given by von Neumann's theorem, depends on both the linear structure and the symplectic form $\omega$ on $\mathbb{V}$.
Specifically, von Neumann’s theorem states that the realizations of a Weyl system on the Hilbert spaces of square-integrable functions on different Lagrangian subspaces of the same symplectic vector space are unitarily related.
This means that every invertible smooth map $\phi$ from $\mathbb{V}$ to itself such that it preserves the given linear structure and symplectic bilinear form $\omega$ on $\mathbb{V}$,  will give rise to a unitary transformation between the von Neumann realization of $(\mathbb{V},\,\omega)$ on the space of square-integrable functions on the Lagrangian subspace $L$ with respect to the Lebesgue measure on it, and the von Neumann realization of $(\mathbb{V},\,\omega)$ on the space of square-integrable functions on the Lagrangian subspace $\phi(L)$  with respect to the Lebesgue measure on it.

This result clearly depends on the fact that we fix the linear structure as well as the symplectic form on $\mathbb{V}$.
Consequently, if alternative linear structures and alternative symplectic forms are given on the same set $\mathbb{V}$, we obtain alternative Weyl systems as well as alternative realizations of Weyl systems in terms of the von Neumann theorem.
This instance is thoroughly investigated in \cite{ErcolessiIbortAlternative}, and will be briefly recalled here by means of a concrete example.

Consider a symplectic vector space $(\mathbb{V},\,\omega)$, where $\mathbb{V} \,=\, \mathbb{R}^2$ with the standard vector space structure, and let $(q,p)$ be Cartesian coordinates adapted to the linear structure of $\mathbb{V}$ in which $\omega$ takes its canonical form.
Next, consider the nonlinear diffeomorphism $\phi$ of $\mathbb{V}$ to itself given by:
\be
(Q,\,P)\,\equiv\,\phi(q,p)\,:=\,(q\,K(|q|),\,p)\,,
\ee
where $K$ is a smooth function, such that $K(|q|) + q \frac{\partial K(|q|)}{\partial q} \, \neq \,0$, in order for this to represent a diffeomorphism.
We may use $(Q,P)$ as Cartesian coordinates adapted to a new linear structure on $\mathbb{V}=\mathbb{R}^{2}$ given by the addition operation $(Q,P)+(Q',P')=(Q+Q',P+P')$ and the scalar multiplication operation $a\cdot(Q,P)=(aQ,aP)$ with $a\in\mathbb{R}$.
This new vector space will be denoted by $\mathbb{V}_{\phi}$.
Note that the vector space structure on $\mathbb{V}_{\phi}$ may be described in the old vector space $\mathbb{V}$ by the following addition and scalar multiplication operations expressed in the old coordinates:
\be
(q,\,p)\,+_\phi \,(q',\,p')\,:=\, \phi^{-1}( \phi(q,\,p) \,+\, \phi(q',\,p'))
\ee
\be
a\,\cdot_\phi \,(q,\,p)\,:=\,\phi^{-1}(a\,\cdot\,\phi(q,\,p))\,,
\ee
where $+$ and $\cdot$ are the addition and scalar multiplication operations on $\mathbb{V}_{\phi}$.
Then, we may take $\omega_{\phi}$ to be the symplectic bilinear form on $\mathbb{V}_{\phi}$ which takes its canonical form with respect to  $(Q,P)$ and define the symplectic vector space $(\mathbb{V}_{\phi},\,\omega_{\phi})$.
Note that the symplectic vector spaces $(\mathbb{V},\,\omega)$ and $(\mathbb{V}_{\phi},\,\omega_{\phi})$ are nonlinearly related by means of $\phi$.

Now, we build the Weyl system associated with $(\mathbb{V},\,\omega)$ by selecting the Lagrangian subspace $\mathcal{L}\,=\, \mathrm{span}\{\,(q,\,0)\,\}$ endowed with the Lebesgue measure $\mathrm{d}\mu=\mathrm{d}q$, so that the Hilbert space of the von Neumann representation is $\mathfrak{L}^2(\mathcal{L},\, \mathrm{d}q)$. 
The operators $U$ and $V$ are then defined by:
\be
\begin{split}
(U(\alpha)\psi)(x) &\,=\, e^{i\alpha q}\psi(x) \\
(V(\beta)\psi)(x) &\,=\, \psi(x\,+\,\beta)
\end{split}
\ee
whose generators turn to be $\hat{x}\,=\,q$ and $\hat{\pi}\,=\,-i\frac{\partial}{\partial q}$ which satisfy the canonical commutation relations on the Hilbert space  $\mathfrak{L}^2(\mathcal{L},\, \mathrm{d}q)$.
From these generators we may build the creation and annihilation operators $\hat{a}^{\dagger}\,=\, \frac{\hat{x} - i \hat{\pi}}{\sqrt{2}}$ and $\hat{a}\,=\, \frac{\hat{x}+i \hat{\pi}}{\sqrt{2}}$, so that the Hilbert space  $\mathfrak{L}^2(\mathcal{L},\, \mathrm{d}q)$ "arises" as the Fock space generated by:
\be
|n\rangle\,=\, \frac{1}{\sqrt{n!}} (\hat{a}^\dag)^n \, |0\rangle
\ee
with $n= 0,1,2,...,n,...$ and where $|0\rangle$ is the vacuum state annihilated by $\hat{a}$.

Similarly, we build the Weyl system associated with $(\mathbb{V}_{\phi},\,\omega_{\phi})$ by selecting the Lagrangian subspace $\mathcal{L}' \,=\, \mathrm{span}\{ \,(Q,\,0) \, \}$ endowed with the Lebesgue measure $\mathrm{d}\mu=\mathrm{d}Q$, so that the Hilbert space is $\mathfrak{L}^2(\mathcal{L}',\,\mathrm{d}Q)$ and the operators $U'$ and $V'$ are given by:
\be
\begin{split}
(U'(\alpha)\psi)(x') &\,=\, e^{i\alpha Q}\psi(x') \\
(V'(\beta)\psi)(x') &\,=\, \psi(x'\,+\,\beta)
\end{split}
\ee
whose generators turn to be $\hat{x}'\,=\, Q$ and $\hat{\pi}'\,=\, -i \frac{\partial}{\partial Q}$  with canonical commutation relations on the Hilbert space $\mathfrak{L}^2(\mathcal{L}',\,\mathrm{d}Q)$.
Again, we may build creation and annihilation operators $\hat{A}^{\dagger}\,=\, \frac{\hat{x}' - i \hat{\pi}'}{\sqrt{2}}$ and $\hat{A}\,=\, \frac{\hat{x}'+i \hat{\pi}'}{\sqrt{2}}$, so that the Hilbert space  $\mathfrak{L}^2(\mathcal{L}',\,\mathrm{d}Q)$ "arises" as the Fock space generated by:
\be
|N\rangle\,=\, \frac{1}{\sqrt{N!}} (\hat{A}^\dag)^n \, |0_{\phi}\rangle
\ee
with $N= 0,1,2,...,n,...$ and where $|0_{\phi}\rangle$ is the vacuum state annihilated by $\hat{A}$.
Note that we may realize the operator $V'(\beta)$ on the Hilbert space $\mathfrak{L}^2(\mathcal{L},\, \mathrm{d}q)$ where it implements translations with respect to the addition operation $+_{\phi}$:
\be
(V'(\beta)\psi)(x) \,=\, \psi(x\,+_{\phi}\,\beta)\,.
\ee

Now, we note that the Lagrangian subspaces $\mathcal{L}$ and $\mathcal{L}'$ coincide because they are the subspaces characterized by $p\,=\,P\,=\, 0$.
However, since the linear structure on $\mathbb{V}$ is nonlinearly related with the linear structure on $\mathbb{V}_{\phi}$, it follows that the Lebesgue measures on $\mathcal{L}$ and $\mathcal{L}'$ are no longer linearly related, and thus square integrable functions with respect to one measure need not be square integrable with respect to the other.
In particular, we may obviously look at $\hat{x}$ and $\hat{\pi}$ as linear operators on both $\mathfrak{L}^2(\mathcal{L},\, \mathrm{d}q)$ and $\mathfrak{L}^2(\mathcal{L}',\, \mathrm{d}Q)$ because $\mathcal{L}$ and $\mathcal{L}'$ are the same subsets of $\mathbb{R}^{2}$, however, it turns out that $\hat{x}$ is self-adjoint on both the Hilbert spaces, while $\hat{\pi}$ is self-adjoint only on  $\mathfrak{L}^2(\mathcal{L},\, \mathrm{d}q)$.
Consequently, the algebra "generated" by the operators $\hat{x},\,\hat{\pi},\,\mathbb{I}$ together with their adjoints on $\mathfrak{L}^2(\mathcal{L},\, \mathrm{d}q)$ is actually a $C^{*}$-algebras, while that generated by $\hat{x},\,\hat{\pi},\,\mathbb{I}$ together with their adjoints on $\mathfrak{L}^2(\mathcal{L}',\, \mathrm{d}Q)$ is not (see \cite{ErcolessiIbortAlternative} for more details).
From this discussion, it should be clear that, in the construction of Weyl systems and their associated von Neumann realizations, it should be explicitely stated the relevant assumption regarding the existence of a specific (and fixed) symplectic vector space structure on $\mathbb{V}$.
Furthermore, it should be clear that, whenever alternative symplectic vector space structures are available at the same time, we may face a particularly rich situation in which nonlinearly related formulations of quantum mechanics are possible.

\section{Conclusions}

The description of a physical system by means of a vector field $\Gamma$ on some carrier manifold $\mathcal{M}$ is something which we should arrive at, rather than to start with.
Experimental data provide us with trajectories on some configuration space, and it is part of the job of a theoretician to extract a differential equation out of them.
As we have argued in section \ref{Sec:Trajectories}, in general it is possible to pass from the experimental trajectories to an implicit differential equation.
This is a subset (hopefully a submanifold) of some carrier space $\mathcal{M}$ which has to be built out of the experimental data.
In this case, obtaining a solution to the inverse problem of the dynamics, that is, finding a Lagrangian or Hamiltonian description of the given physical system, is particularly difficult.
Some comments on this situation are given in appendix \ref{App:InverseProblemImplicit}.
However, in some cases it is possible to pass from the experimental trajectories to an explicit differential equations the solutions of which are (appropriate lifts of) the trajectories we started with, and we obtain a vector field $\Gamma$ on the carrier manifold $\mathcal{M}$.

After explaining the main points necessary to the construction of $\Gamma$ from experimental trajectories, we passed, in sections \ref{sec: lagrangian} and \ref{sec: hamiltonian}, to analyse the process of "tensorialization" of the Lagrangian and Hamiltonian description of the physical systems associated with $\Gamma$.
In this way, it is possible to single out the qualitative features that characterize a given picture of Classical Mechanics, and the possibility of obtaining alternative Lagrangian or Hamiltonian descriptions for the same physical systems is clearly enlightened.
In the Lagrangian picture, the carrier manifold $\mathcal{M}$ turns out to be diffeomorphic to the tangent bundle $\textbf{T}\mathcal{Q}$ of the configuration space $\mathcal{Q}$, and, once the tangent bundle structure is fixed, alternative Lagrangian descriptions are usually obtained by means of the construction (if possible) of alternative Lagrangian functions for the same vector field $\Gamma$ on $\mathcal{M}\cong T\mathcal{Q}$ (\cite{MarmoTangent}).
However, once the tangent bundle structure of $\mathcal{M}$ is "tensorialized" in the couple $(S,\,\Delta)$ as it is recalled in section \ref{sec: lagrangian},  we pointed out that it is also possible to change the tangent bundle structure of $\mathcal{M}$ so that the same vector field $\Gamma$ becomes a second order vector field for an alternative tangent bundle structure, and thus the possibility of a Lagrangian description of $\Gamma$ with respect to this alternative tangent bundle structure has to be investigated.
This instance can also be read from the opposite point of view.
Specifically, we may look for a tangent bundle structure on $\mathcal{M}$ (if possible) in which $\Gamma$ is a second order vector field admitting of a Lagrangian description.
Furthermore, in some cases it could be necessary to reparametrize $\Gamma$ in such a way that it becomes a complete vector field (e.g., the Kepler problem), and this reparametrization would in general make the reparametrized vector field no longer second order with respect to the "old" tangent bundle structure so that a Lagrangian description for the reparametrized vector field is not possible.
However, we may ask if there is an alternative tangent bundle structure on $\mathcal{M}$ with respect to which the reparametrized vector field is a second order vector field so that search for a Lagrangian description for the reparametrized vector field in the "new" tangent bundle structure is meaningful. 
At the end of section \ref{sec: lagrangian} we provided some simple examples where this program can be succesfully followed, and we plan to take on a more systematic analysis of this instance in future works.

In section \ref{sec: hamiltonian}, we applied the "tensorialization" process to the Hamiltonian picture of dynamics.
In this case, we showed how the diffeomorphism invariance of the tensorial description allows us to obtain alternative Hamiltonian descriptions associated with symmetries and constants of the motion.
Specifically, let $\Gamma$ be the dynamical vector field , which is assumed to be the Hamiltonian vector field of some Hamiltonian function $\mathscr{H}$ with respect to the symplectic form $\omega$.
If $\phi\colon\mathcal{M}\rightarrow\mathcal{M}$ is a diffeomorphism which is a symmetry for the dynamics ($\phi^{*}\Gamma=\Gamma$), we may form the symplectic form $\omega_{\phi}=\phi^{*}\omega$ and the $(1-1)$ tensor field $T_{\phi}=\Lambda\circ\omega_{\phi}$.
Clearly, both $\omega_{\phi}$ and $T_{\phi}$ are invariant with respect to $\Gamma$.
Then, if $f$ is a constant of the motion for the dynamics ($\Gamma(f)=0$), we have that $\omega_{f}=\mathrm{d}\,\mathrm{d}_{T_{\phi}}\,f=\mathrm{d}\, T_{\phi}(\mathrm{d}f)$ is a two-form which is invariant with respect to the dynamical vector field $\Gamma$.
When $\omega_{f}$ is non-degenerate, it provides us with an alternative Hamiltonian description for $\Gamma$ as the locally Hamiltonian vector field for the closed one-form $i_{\Gamma}\omega_{f}$.

In section \ref{sec: quantum}, we recalled the deep connection between alternative symplectic vector space structures and alternative Weyl systems leading to alternative quantum descriptions and alternative commutation relations.
Indeed, a Weyl system is a map from a symplectic vector space $(\mathbb{V}, \omega)$ to the group of unitary operators on some Hilbert space satisfying additional properties (see section \ref{sec: quantum} for details).
Consequently, the existence of alternative linear structures and alternative symplectic structures on $\mathbb{V}$ gives rise to an alternative structure of symplectic vector space on $\mathbb{V}$ and to an alternative realization of the Weyl system by means of von Neumann's theorem.
According to the discussion in section \ref{sec: hamiltonian}, alternative symplectic structures may be "dynamically" obtained starting with a (linear) vector field on $\mathbb{V}$ thus showing how to pass from classical-like trajectories to the quantum commutation relations encoded in the infinitesimal generators of the Weyl system.
Furthermore, the coexistence of nonlinearly related alternative symplectic vector space structures  implies the coexistence of nonlinearly related alternative formulations of quantum mechanics, together with the coexistence of nonlinearly related alternative procedures of second quantization (\cite{ErcolessiCanonical}).

\section*{Acknowledgments}
GM would like to thank partial financial support provided by the Santander/UC3M Excellence  Chair Program 2019/2020.

\appendix

\section{Inverse problem for implicit differential equations} \label{App:InverseProblemImplicit}

In section \ref{Sec:Trajectories} we saw how, in general, experimental data lead the theoretician to build a submanifold of $\textbf{TT}\mathcal{Q}$, that is, an implicit differential equation on (a submanifold of) the tangent bundle $\textbf{T}\mathcal{Q}$. 
Within this context the inverse problem is much more complicated to address than it is in the explicit case.
Essentially, this is due to the fact that the \emph{Euler-Lagrange equations} are formulated by means of an implicit differential equation on $\textbf{T}^*\mathcal{Q}$ rather than by means of an implicit differential equation on  $\textbf{T}\mathcal{Q}$, even though the Lagrangian function is defined on the tangent bundle $\textbf{T}\mathcal{Q}$.
Indeed, given the Lagrangian function $\lag$, we have $\mathrm{d}\lag\colon \textbf{T}\mathcal{Q}\rightarrow\,\textbf{T}^{*}\textbf{T}\mathcal{Q}$, and thus $\mathrm{d}\lag(\textbf{T}\mathcal{Q})$ is a submanifold of $\textbf{T}^{*}\textbf{T}\mathcal{Q}$.
By means of the inverse of the canonical Tulczyjew isomorphism $\tau\colon \textbf{TT}^*\mathcal{Q}\rightarrow \textbf{T}^*\textbf{T}\mathcal{Q}$ (\cite{TulczyjewTriples}), we obtain a submanifold of $\textbf{TT}^*\mathcal{Q}$, i.e., an implicit differential equation on $\textbf{T}^*\mathcal{Q}$. Explicitly, we have
\be
\mathrm{d}\lag \;\; |\;\; \textbf{T}\mathcal{Q} \to \textbf{T}^*\textbf{T}\mathcal{Q} \;\; : \;\; (q^i,\, v^i) \mapsto \left(q^i,\,v^i,\, \frac{\partial \lag}{\partial q^i},\, \frac{\partial \lag}{\partial v^i} \right) \,,
\ee
while the Tulczyjew isomorphism between $\textbf{TT}^*\mathcal{Q}$ and $\textbf{T}^*\textbf{T}\mathcal{Q}$ is defined by 
\be
\tau \;\;|\;\; \textbf{TT}^*\mathcal{Q} \to \textbf{T}^*\textbf{T}\mathcal{Q} \;\;:\;\; (q^i,\,p_i,\,v_q^i,\, {v_p}_i) \mapsto (q^i,\,v_q^i,\,{v_p}_i,\,p_i)
\ee
(see \cite{TulczyjewTriples} section $3$). This is, in fact, a \textsc{symplectomorphism} mapping the canonical symplectic form over $\textbf{T}^*\textbf{T}\mathcal{Q}$, i.e., $\mathrm{d}q^i \wedge \mathrm{d}{p_q}_i + \mathrm{d}v^i \wedge \mathrm{d}{p_v}_i$, into the canonical symplectic form over $\textbf{TT}^*\mathcal{Q}$, i. e., $\mathrm{d}v_q^i \wedge \mathrm{d}{p_v}_i + \mathrm{d}q^i \wedge \mathrm{d}{p_q}_i$. By composing $\mathrm{d}\lag$ and $\tau^{-1}$ we  obtain a submanifold of $\textbf{T}^*\textbf{T}\mathcal{Q}$, say $\Sigma$, given by
\be \label{Eq: submanifold lagrange equations}
 \Sigma\,:=\,\left(\tau^{-1} \circ \mathrm{d}\lag\right) (\textbf{T}\mathcal{Q})  \,=\,\left\{ \, (q^i,p_i,\,{v_q}^i ,\, {v_p}_i)\in \textbf{TT}^*\mathcal{Q} \;\;|\;\; p_i \,=\, \frac{\partial \lag}{\partial v^i}\,, \;\; {v_p}_i \,=\, \frac{\partial \lag}{\partial q^i} \, \right\} \,.
\ee
Writing $i_{\Sigma}$ for the canonical immersion of $\Sigma$ in $\textbf{TT}^*\mathcal{Q}$, it follows that $\Sigma$ is a Lagrangian (or simply isotropic if L is not regular) submanifold of $\textbf{TT}^*\mathcal{Q}$ because
\be
i_{\Sigma}^{*}\left(\mathrm{d}v_q^i \wedge \mathrm{d}{p_v}_i + \mathrm{d}q^i \wedge \mathrm{d}{p_q}_i \,\right) \, = \, \mathrm{d}v_q^i \wedge\,\mathrm{d} \left( \frac{\partial \lag}{\partial v^i} \right) + \mathrm{d}q^i \wedge \mathrm{d}\left( \frac{\partial \lag}{\partial q^i} \right) \,=\, 0
\ee
since $\lag$ depends only on $(q^i,\,{v_q}^i)$.
Note that, with the prescription that ${v_p}_i \,=\, \frac{d}{dt}p_i$, it immediately follows that $\Sigma$ as defined in equation \eqref{Eq: submanifold lagrange equations} is the submanifold of $\textbf{TT}^*\mathcal{Q}$ on which the \textsc{Euler-Lagrange equations}
\be
\frac{d}{dt}\frac{\partial \lag}{\partial v^i} \,=\, \frac{\partial \lag}{\partial q^i}
\ee
are identically satisfied.

\begin{remark}
A similar construction is possible for the Hamiltonian case where it clearly emerges that the submanifold of $\textbf{TT}^*\mathcal{Q}$ one obtains is the graph of a vector field, that is, the equations are always explicit ones. Consider the cotangent bundle $\textbf{T}^*\mathcal{Q}$ of the configuration space $\mathcal{Q}$ and a Hamiltonian function $\mathscr{H} \;\;|\;\; \textbf{T}^*\mathcal{Q} \to \mathbb{R}$. Via its differential
\be
\mathrm{d}\mathscr{H}\;\;|\;\;\textbf{T}^*\mathcal{Q} \to \textbf{T}^*\textbf{T}^*\mathcal{Q} \,,
\ee
we can define a submanifold $\Sigma$ of $\textbf{T}^*\textbf{T}^*\mathcal{Q}$ by setting $\Sigma:=\mathrm{d}\mathscr{H}(\textbf{T}^*\mathcal{Q})$.
Differently from the Lagrangian case, we have that $\textbf{T}^*\textbf{T}^*\mathcal{Q}$ is isomorphic to $\textbf{TT}^*\mathcal{Q}$  because the Poisson structure $\Lambda$  associated with the canonical symplectic structure over $\textbf{T}^*\mathcal{Q}$ define an isomorphism between differential forms and vector fields on $\textbf{T}^*\mathcal{Q}$.
With an evident abuse of notation, we denote this isomorphism with $\Lambda$.
By means of $\Lambda$, we obtain the implicit differential equation $\Lambda(\Sigma)$ on $\textbf{TT}^*\mathcal{Q}$.
Denoting by $X_{\mathscr{H}}$ the Hamiltonian vector field associated with $\mathscr{H}$ by means of the Poisson tensor $\Lambda$, that is, $X_{\mathscr{H}}=\Lambda(\mathrm{d}\mathscr{H})$, the following diagram may be defined:
\be
\begin{tikzcd}
& \textbf{TT}^*\mathcal{Q} & & \arrow{ll}[swap]{\Lambda} \textbf{T}^*\textbf{T}^*\mathcal{Q} \\
&  & \textbf{T}^*\mathcal{Q}  \arrow{ul}{X_{\mathscr{H}}} \arrow{ur}[swap]{\mathrm{d}\mathscr{H}}\,, &
\end{tikzcd}
\ee
and it follows that $\Lambda(\Sigma)$ is precisely $X_{\mathscr{H}}(\textbf{T}^*\mathcal{Q})$. 
Consequently, being $\Lambda(\Sigma)$ the image of $\textbf{T}^*\mathcal{Q}$ through a vector field, it emerges that the dynamics is always an explicit one in the Hamiltonian case, as it is also clear from the standard form of the Hamilton equations.
\end{remark}

We can say that \textsc{Euler-Lagrange equations} force us to work with a Lagrangian submanifold of $\textbf{TT}^*\mathcal{Q}$, while, on the one hand, we saw in section \ref{Sec:Trajectories}  how experimental data would naturally lead us to build a submanifold of $\textbf{TT}\mathcal{Q}$.
Consequently, the following question is unavoidable: \emph{how can these two seemingly uncompatible instances be related?} 
The essential difficulty is due to the absence of a natural, ``pre-existing'', symplectic structure on $\textbf{T}\textbf{T} \mathcal{Q}$.
To be able to formulate the inverse problem for the submanifold of $\textbf{TT}\mathcal{Q}$ we construct out of the trajectories on $\mathcal{Q}$, we would need a map:
\begin{equation}
\phi \;\; : \;\; \textbf{T}\mathcal{Q} \to \textbf{T}^*\mathcal{Q}
\end{equation}
so that we would be able to map the submanifold of $\textbf{TT}\mathcal{Q}$, that we constructed out of trajectories, onto a submanifold of $\textbf{T}\textbf{T}^*\mathcal{Q}$ by means of the tangent map $T\phi$:
\begin{equation}
\begin{tikzcd}
\textbf{TT}\mathcal{Q} \arrow{d}[swap]{\pi_{\textbf{T}}} \arrow[rr, "T\phi"]  &    & \textbf{TT}^*\mathcal{Q} \arrow[d, "\pi_{\textbf{T}^*}"] \\
\textbf{T}\mathcal{Q} \arrow[rr, "\phi"] \arrow[dr] & & \textbf{T}^*\mathcal{Q} \arrow[dl] \\
 & \mathcal{Q} &
\end{tikzcd}
\end{equation}

It may happen that the Lagrangian function itself could provide us with the map $\phi$ by means of the \textsc{fiber derivative} 
\be
\mathscr{F}\lag \;|\;\; \textbf{T}\mathcal{Q} \to \textbf{T}^*\mathcal{Q} \;:\;\; (q^i, \, v^i) \mapsto \pi_{\textbf{T}^*}\, \circ \, \tau \, \circ \, \mathrm{d}\lag = \left(q^i, \frac{\partial \lag}{\partial v^i} \right)\,.
\ee
This map coincides with the map defined by the following diagram:

\begin{equation}
\begin{tikzcd}
\textbf{TT}\mathcal{Q} \arrow{d}[swap]{\pi_{\textbf{T}}} \arrow[rr, bend left, "T\mathscr{F}\lag"]  &  \textbf{T}^*\textbf{T}\mathcal{Q} \arrow[r, rightharpoonup, shift left, "\tau"] \arrow[r, leftharpoondown, shift right]  & \textbf{TT}^*\mathcal{Q} \arrow[d, "\pi_{\textbf{T}^*}"] \\
\textbf{T}\mathcal{Q} \arrow[ur, "\mathrm{d}\lag"] \arrow[rr, "\mathscr{F}\lag"] \arrow[dr] & & \textbf{T}^*\mathcal{Q} \arrow[dl] \\
 & \mathcal{Q} &
\end{tikzcd}
\end{equation}

By means of the fiber derivative, the canonical symplectic structure $\Omega = \mathrm{d}p_i \wedge \mathrm{d}q^i$ and its potential $\theta = p_i \mathrm{d}q^i$  on $\textbf{T}^*\mathcal{Q}$ can be pulled-back on $\textbf{T}\mathcal{Q}$ to obtain
\begin{equation}
\begin{split}
&\theta_{\lag} = (\mathscr{F}\lag)^* \theta = \partial^v_i \lag  \mathrm{d}q^i \, \\  \Omega_{\lag} = (\mathscr{F}\lag)^*&\Omega = \frac{\partial^2 \lag}{\partial q^i \partial v^j} \mathrm{d}q^i \wedge \mathrm{d}q^j + \frac{\partial^2 \lag}{\partial v^i \partial v^j} \mathrm{d}v^i \wedge \mathrm{d}q^j\,.
\end{split}
\end{equation}
By means of its tangent map, $T\mathscr{F}\lag$, the symplectic structure on $\textbf{TT}^*\mathcal{Q}$ and its potential, $\dot{\Omega}$ and $\dot{\theta}$, can be pulled-back on $\textbf{TT}\mathcal{Q}$:
\be
\begin{split}
\dot{\theta}_{\lag} = (T\mathscr{F}\lag)^* \dot{\theta} \\ \dot{\Omega}_{\lag} = (T\mathscr{F}\lag)^*\dot{\Omega}\,.
\end{split}
\ee
Note that, in general, $\Omega_{\lag}$ and $\dot{\Omega}_{\lag}$ are no longer symplectic form because they may present a kernel which depends on $\lag$.

In conclusion, the Lagrangian plays a double role within the formulation of the inverse problem for implicit differential equations. First, it defines a Lagrangian submanifold $\Sigma$ of $\textbf{TT}^*\mathcal{Q}$, which represents the Lagrangian formulation of the dynamics. Second, it allows for the definition of a fiber derivative $\mathscr{F}\lag$ which, if suitable regularity conditions on $\lag$ are satisfied, would make it possible to impose that the pre-image of $\Sigma$ through $T\mathscr{F}\lag$ coincide with the submanifold of $\textbf{TT}\mathcal{Q}$ on which the experimental data naturally live.
See also section $2.1$ in \cite{DeDiegoInverse} for another discussion about the inverse problem for implicit differential equations.


\begin{thebibliography}{10}

\bibitem{Aristotele}
Aristotele.
\newblock {\em {P}hysics. {E}nglish translation by {P}hilip {W}icksteed and
  {F}rancis {C}ornford}.
\newblock Heinemann, London 1957.

\bibitem{MarmoSaletan}
G.~Marmo, E.~J. Saletan, A.~Simoni, and B.~Vitale.
\newblock {\em Dynamical systems: a differential geometric approach to symmetry
  and reduction}.
\newblock Wiley-Interscience, New York, 1985.

\bibitem{DavanzoKepler}
A.~D'Avanzo and G.~Marmo.
\newblock Reduction and unfolding: the {K}epler problem.
\newblock {\em International Journal of Geometric Methods in Modern Physics},
  2(1):83,109, 2005.
  
\bibitem{DavanzoQuantum}
A.~D'Avanzo, G.~Marmo and A.~Valentino.
\newblock Reduction and unfolding for quantum systems: the {H}ydrogen atom.
\newblock {\em International Journal of Geometric Methods in Modern Physics},
  2(6):1043,1062, 2005.

\bibitem{MankoFoscillator}
V.~I. Man'ko, G.~Marmo, E.~C.~G. Sudarshan, and F.~Zaccaria.
\newblock f-oscillators and nonlinear coherent states.
\newblock {\em Physica Scripta}, 55(5):528, 1997.

\bibitem{MasterThesis}
L.~Schiavone.
\newblock From trajectories to commutation relations.
\newblock Master's thesis, Università degli studi di Napoli Federico II, 2018.

\bibitem{MarmoMendellaTulczyjewImplicit}
G.~Marmo, G.~Mendella, and W.~Tulczyjew.
\newblock Constrained {H}amiltonian system a implicit differential equations.
\newblock {\em Journal of Physics A}, 30:277,293, 1997.

\bibitem{KrupkaGlobal}
D.~Krupka.
\newblock {\em Introduction to {G}lobal {V}ariational {G}eometry}.
\newblock Atlantis Press, Paris 2015.

\bibitem{KrupkovaOVE}
O.~Krupkova.
\newblock {\em The geometry of ordinary variational equations}.
\newblock Springer, Berlin Heidelberg, 1997.

\bibitem{KrupkovaInverse}
O.~Krupkova.
\newblock {\em Second order ordinary differential equations on Jet Bundles and the Inverse Problem of the Calculus of Variations}.
\newblock {\em Handbook of Global Analysis}, Elsevier, New York, 1997.

\bibitem{SardanashvilyNewMethods}
G.~Giachetta, L.~Mangiarotti, and J.~Sardanashvily.
\newblock {\em New {L}agrangian and {H}amiltonian methods in {F}ield {T}heory}.
\newblock World Scientific, Singapore, 1996.

\bibitem{GeometryFromDynamics}
J.~Carinena, G.~Marmo, A.~Ibort, and G.~Morandi.
\newblock {\em Geometry from {D}ynamics, {C}lassical and {Q}uantum}.
\newblock Springer, Heidelberg, 2015.

\bibitem{Santilli}
R.~M. Santilli.
\newblock {\em Foundations of {T}heoretical {M}echanics {I}: the {I}nverse
  problem in {N}ewtonian {M}echanics}.
\newblock Springer-verlag, New York, 1978.

\bibitem{MarmoTangent}
G.~Morandi, C.~Ferrario, G.~Lo~Vecchio, G.~Marmo, and C.~Rubano.
\newblock The inverse problem in the {C}alculus of {V}ariations and the geometry
  of the tangent bundle.
\newblock {\em Physics Report}, 188(3 \& 4):147,284, 1990.

\bibitem{CrampinInverse}
M.~Crampin.
\newblock {O}n the differential geometry of the {E}uler-{L}agrange equations,
  and the inverse problem of {L}agrangian dynamics.
\newblock {\em Journal of Physics A: Math. Gen.}, 14:2567,2575, 1981.

\bibitem{MarmoGiordanoRubanoInverseHamiltonian}
M.~Giordano, G.~Marmo, and G.~Rubano.
\newblock The inverse problem in the {H}amiltonian formalism: integrability of
  linear {H}amiltonian fields.
\newblock {\em Inverse problem}, 9:443,467, 1993.

\bibitem{MarmoTensorialTangentBundle}
S.~De~Filippo, G.~Landi, G.~Marmo, and G.~Vilasi.
\newblock Tensor fields defining a tangent bundle structure.
\newblock {\em Annales de l'I.H.P.}, A(50-2):205,218, 1989.

\bibitem{abraham_marsden_ratiu-manifolds_tensor_analysis_and_applications}
R.~Abraham, J.~E. Marsden, and T.~Ratiu.
\newblock {\em {Manifolds, tensor analysis, and applications}}.
\newblock Springer-Verlag, New York, 3rd edition, 2012.

\bibitem{PichereauPoisson}
C.~Laurent-Gengoux, A.~Pichereau, and P.~Vanhaecke.
\newblock {\em Poisson structures}.
\newblock Springer, Berlin Heidelberg, 2013.

\bibitem{SmirnovMagriCompatibility}
R.~G. Smirnov.
\newblock Magri–{M}orosi–{G}el’fand–{D}orfman’s bi-{H}amiltonian
  constructions in the action-angle variables.
\newblock {\em Journal of Mathematical Physics}, 38(12):6444,6453, 1997.


\bibitem{ErcolessiIbortAlternative}
E.~Ercolessi, A.~Ibort, G.~Marmo, and G.~Morandi.
\newblock Alternative linear structures for classical and quantum systems.
\newblock {\em International Journal of Modern Physics A}, 22(18):3039,3064,
  2007.
  
\bibitem{ErcolessiCanonical}
E.~Ercolessi, G.~Marmo, and G.~Morandi.
\newblock From equations of motion to canonical commutation relations.
\newblock {\em Rivista del Nuovo Cimento}, 33: 401,590, 2010.

\bibitem{TulczyjewTriples}
J.~Grabowski, M.~Kùs, G.~Marmo, and T.~Shulman.
\newblock Geometry of quantum dynamics in infinite-dimensional {H}ilbert space.
\newblock {\em Journal of Physics A: Math. Theor.}, 51, 2018.

\bibitem{DeDiegoConstrainedInverse}
M.~Barbero-Linan, M.~Farre~Puiggal\'{i} and D.~M. De~Diego.
\newblock Isotropic submanifolds and the inverse problem for mechanical constrained systems.
\newblock {\em Journal of Physics A: Mathematical and Theoretical}, 48(4),
  2015.

\bibitem{DeDiegoInverse}
M.~Barbero-Linan, M.~Farre~Puiggal\'{i}, S.~Ferraro, and D.~M. De~Diego.
\newblock The inverse problem of the calculus of variations for discrete
  systems.
\newblock {\em Journal of Physics A: Mathematical and Theoretical}, 51(18),
  2018.

\end{thebibliography}
\end{document}